# Active Shape-Morphing Elastomeric Colloids in Short-Pitch Cholesteric Liquid Crystals

Julian S. Evans,[1] Yaoran Sun,[1,2] Bohdan Senyuk,[1] Patrick Keller,[3,4] Victor M. Pergamenshchik,[6] Taewoo Lee,[1] and Ivan I. Smalyukh[1,5,7,*]

[1]Department of Physics, University of Colorado, Boulder, Colorado 80309, USA
[2]Centre for Optical and Electromagnetic Research, Zhejiang University, Hangzhou 310058, People's Republic of China
[3]Institut Curie, Centre de Recherche, CNRS UMR 168, Université Pierre et Marie Curie, 75248 Paris cedex 05, France
[4]Department of Chemistry and Biochemistry and Liquid Crystal Materials Research Center, University of Colorado, Boulder, Colorado 80309, USA
[5]Material Science and Engineering Program, Department of Electrical, Computer, & Energy Engineering, and Liquid Crystal Materials Research Center, University of Colorado, Boulder, Colorado 80309, USA
[6]Institute of Physics, Prospect Nauky 46, Kyiv, 03039, Ukraine
[7]Renewable and Sustainable Energy Institute, National Renewable Energy Laboratory and University of Colorado, Boulder, Colorado 80309, USA

*Email: ivan.smalyukh@colorado.edu

**Abstract**

Active elastomeric liquid crystal particles with initial cylindrical shapes are obtained by means of soft lithography and polymerization in a strong magnetic field. Gold nanocrystals infiltrated into these particles mediate energy transfer from laser light to heat, so that the inherent coupling between the temperature-dependent order and shape allows for dynamic morphing of these particles and well-controlled stable shapes. Continuous changes of particle shapes are followed by their spontaneous realignment and transformations of director structures in the surrounding cholesteric host, as well as locomotion in the case of a nonreciprocal shape morphing. These findings bridge the fields of liquid crystal solids and active colloids, may enable shape-controlled self-assembly of adaptive composites and light-driven micromachines, and can be understood by employing simple symmetry considerations along with electrostatic analogies.

Dispersions of particles in liquid crystals (LCs) attract a constantly growing interest and exhibit many fascinating phenomena, ranging from particle-induced topological defects to novel types of elastic interactions [1–3]. Structures of LC molecular alignment around particles, typically described by a director **n** pointing along the local average molecular orientation, give rise to elasticity mediated colloidal self-assembly not encountered in isotropic fluid hosts. Within a far-field approximation, even spherical particles with strong boundary conditions induce dipolar or quadrupolar **n(r)** distortions arising due to the anisotropic nature of the host medium and particle-induced satellite defects, such as point defects and disclination loops [1,2]. Although many methods for self-assembly of anisotropic particles in isotropic solvents have been introduced [4], LC hosts bring a number of unique capabilities, such as the well-defined long-range alignment of anisotropic particles with respect to **n** controlled by external fields [3,5–10]. The shape of the particles dictates colloidal self-assembly [3,7–15] and may enable practical utility of these fascinating LC-colloidal systems. Although complex-shaped particles can be fabricated using photolithography [3,6], two-photon polymerization [10], nanocrystal growth [8], and wet chemical synthesis [7,9], these approaches provide nematic dispersions of nonresponsive particles with no means for altering their shape, alignment, locomotion, and self-assembly [3,5–8].

In this Letter, we describe an optically reconfigurable double-LC colloidal system that allows for an unprecedented optical control of particle shapes and motion while being dispersed in a LC host. These active particles are made from a nematic LC elastomer (LCE) [16–18] doped with gold nanocrystals to enable photothermal energy transfer and are morphed into low-symmetry configurations via laser-directed bending and aspect ratio control. We explore director structures in the LC host induced by these active colloids and describe them using electrostatic analogies. Finally, we discuss potential applications in reconfigurable self-assembly and controlled dynamics of active colloids in structured fluid hosts.

Cylindrical particles were obtained using replica molding and polymerization in a strong magnetic field of about 1 T that aligns **n** of the LCE along the cylinder axis [16,17,19]. Particles with diameter×length dimensions of 2×4, 17×70, and 20×100 μm were then passively infiltrated with 2 nm gold nanocrystals (obtained using wet synthesis described in Ref. [20]) by letting them soak in the dispersion of nanocrystals in toluene for 24 h [16]. LCE microparticles with embedded gold nanocrystals were then redispersed in an aqueous cholesteric LC (CLC) solution of hydroxypropylcellulose (HPC, at 55 wt %) [21,22]. The sample was sheared to a partially unwound state which then relaxed to a ground-state CLC configuration with a uniform helical axis perpendicular to the shear direction, as also observed in other studies of HPC-based LCs [21–25]. The equilibrium pitch $p \approx 350$ nm [21–25] and "layer" periodicity [26] $p/2 \approx 175$ nm allow for optical exploration of elastic distortions due to colloidal particles much larger than $p/2$. The LC cells were constructed from 170 μm thick glass plates optimized for high-resolution imaging and laser manipulation when using microscope objectives with high numerical aperture. After shearing, the glass plates were sealed together using epoxy to form cells of thickness ranging from 100 μm to 2 mm.

To control particle shapes, we used laser tweezers [27–29] consisting of a two-axis scanning-mirror head (XLRB2, Nutfield Technology) and a continuous wave Ytterbium-doped fiber laser (wavelength $\lambda = 1064$ nm, IPG Photonics). The tweezers setup was integrated with an optical microscope BX-51 (Olympus). LCE particles in the LC were manipulated via steering a focused beam and monitored by polarizing optical microscopy (POM) using microscope objectives with magnifications 10×, 50×, and 100× and numerical aperture within 0.25–1.4. Linear polarization of the laser beam was controlled by rotating a $\lambda/2$ wave plate. LCE particles were also studied using broadband coherent anti-Stokes Raman scattering polarizing microscopy (CARS-PM) [30,31] via probing the polarized CARS-PM signals from the aromatic C=C stretching within 1400–1600 cm$^{-1}$. Broadband Stokes (obtained by means of a photonic crystal fiber) and 780 nm pump-probe femtosecond pulses were used for excitation. The CARS-PM signal at around 694 nm was detected using a band pass filter with center wavelength at 700 nm and 13 nm bandwidth [31]. POM and CARS-PM [Figs. 1(a)–1(c)] revealed weakly undulating $\mathbf{n}(\mathbf{r})$ of LCE with an average orientation along the cylinder [Figs. 1(d) and 1(e)] [16].

Unlike LCE particles, the LC host was designed to be insensitive to laser beams: (a) optical Fréedericksz transition in this system does not occur up to powers of 1 W (much higher than powers <200 mW used in experiments); (b) polymer solutions of HPC are practically insensitive to photothermal effects, as long as solvent evaporation is avoided; (c) gold nanocrystals uniformly distribute within the LCE but are immiscible with the aqueous solution of HPC, so that they do not diffuse into the surrounding host; (d) having entropic origin, tangentially degenerate surface anchoring of the HPC-based LC is insensitive to the scanned beams. This allows for morphing particle shapes without direct coupling between $\mathbf{n}(\mathbf{r})$ of the

HPC-based host and laser light [Figs. 1(d)–1(l)] as well as POM imaging of the ensuing structural changes. Since the lamellar periodicity $p/2 \approx 175$ nm [21–25] of the HPC cholesteric in the ground state is smaller than the resolution of an optical microscope ($\approx$300 nm or worse, depending on the objective), acetic acid is used as a solvent instead of water to increase $p$ up to 3 μm and directly observe cholesteric layers in POM, as shown in the inset of Fig. 1(l). POM textures obtained without and with a λ-retardation plate also reveal the spatial variations of the helical axis $\chi(\mathbf{r})$ structures.

Unidirectional laser scanning controls the particle shape via photothermal heating mediated by LCE-entrapped gold nanocrystals uniformly distributed within the particle bulk (Fig. 1) [32–35]. LCE microcylinders bend toward the beam scanning direction typically chosen to be orthogonal or tilted with respect to the cylinder axis [Figs. 1(d)–1(l), 2, and 3] [16], giving rise to a number of low-symmetry LCE particles. Bidirectional scanning of a laser beam along the cylinder axis allows for both reversible and irreversible modification of the diameter-length ratio, depending on whether the power of the scanning beam is reduced gradually or switched off abruptly (Fig. 4). POM imaging with and without a λ plate reveals layered structures and $\chi(\mathbf{r})$ of the surrounding LC that follow laser-guided changes of colloidal shapes (Figs. 2–4), which is due to the surface anchoring of $\chi(\mathbf{r})$ at the LC-LCE interface [inset of Fig. 1(l)]. In addition to smooth deformations of layers at relatively weak bending of pillars, POM textures also provide evidence for the presence of dislocations [Fig. 1(l)] that appear to assist in preserving layer equidistance while accommodating boundary conditions on optically controlled LCE particles.

Particle-induced perturbations of a homogeneous ground-state director $\mathbf{n}_0$ in nematic LCs have the form of a transverse two-component vector $\delta\mathbf{n}$ [36]. Along with symmetry considerations, this brings about an expansion into multipole series and electrostatic analogies between long-distance colloidal interactions and that of multipoles in electrostatics [1–11,36]. The ground state in a CLC is strongly twisted and a direct parallel with the homogeneous nematic ground state is impossible because homogeneity of the ground state is a prerequisite of the symmetry-based considerations [11]. However, a similar approach for cholesterics is prompted by the de Gennes–Lubensky model [36], which describes a short-pitch CLC in terms of the director $\chi$ normal to cholesteric layers. The model assumes that distortions occur over length scales much larger than $p$, which is the case of distortions due to our particles of size $\gg p$. The ground state $\chi_0$ of the $\chi$ director is homogeneous and has the same symmetry $D_{\infty h}$ as that of

$\mathbf{n}_0$, and its small distortion is a two-component transverse vector similar to $\delta\mathbf{n}$. Hence allowed symmetries of $\chi$ in CLCs are similar to those of $\mathbf{n}$ in nematics and this approach can be used at distances much larger than the particle size. At the same time, we note that the short distance $\chi(\mathbf{r})$ and $\mathbf{n}(\mathbf{r})$ distortions around particles and the ensuing colloidal assemblies can be fully described only numerically [37].

LCE particle's surface induces a normal alignment of $\chi$ and tangential alignment of the CLC layers [inset of Fig. 1(l)]. The unperturbed axis of a LCE microcylinder is normal to $\chi_0$ and parallel to both the CLC layers and cell substrates. $\chi(\mathbf{r})$ around the cylinder is of the quadrupolar type [Figs. 1(g) and 2(e)] with three mirror symmetry planes: the two planes passing through the cylinder axis, one along $\chi_0$ and the other normal to $\chi_0$, and the plane normal to the cylinder axis. Optical morphing removes some or all of these symmetry planes, thus creating different types of elastic multipoles (Figs. 1–3). One can also vary the strength of distortions without altering the symmetry of ensuing elastic configurations by changing the aspect ratio of particles (Fig. 4) or their fragments [Fig. 1(l)]. Using laser scanning, we can tune the strength and geometrical distribution of elastic multipoles by controlling angles between the bent fragments of the LCE pillar, their diameters, and lengths (Figs. 3 and 4).

Experimental dipolar V- and U-shaped particles with apex along $\chi_0$ shown in Figs. 1(h)–1(j), 2(a)–2(d), and 3(a)–3(c) and zigzag-shaped particles in the right-side parts of Figs. 3(a) and 3(b) induce same-symmetry $\chi(\mathbf{r})$ shown in Figs. 2(f) and 2(g) and Fig. 3(c). Allowed symmetry elements of these $\chi(\mathbf{r})$ structures are those of the group $C_{2v}$: a rotation by an angle $\pi$ about the vertical axis passing through the apex of the V, and a vertical mirror plane lying in the plane of the particle's V and passing through the vertical axis, as further illustrated by sketches of $\chi(\mathbf{r})$ in the vertical and horizontal planes in the supplemental Fig. S1a [38]. The particle shown in Fig. 1(k) has apex of V pointing along layers and normal to $\chi_0$. The symmetry elements of $\chi(\mathbf{r})$ are those of the group $D_{1h}$ and include a vertical mirror plane lying in the particle's plane, a horizontal mirror plane normal to it and passing through the apex of the V, and a $\pi$ rotation about the horizontal axis passing through the apex, as illustrated in the supplemental Fig. S2a [38]. Thus, these dipolar colloids in a CLC are described by point groups $C_{2v}$ or $D_{1h}$, depending on the angle the apex direction makes with $\chi_0$. By continuous morphing particle shape, we can further lower the symmetry of $\chi(\mathbf{r})$ around LCE particles by, for example, removing symmetry elements of $C_{2v}$. Figure 3(d) shows several examples of particles with just one vertical mirror plane coinciding with the plane of the image (symmetry group $C_{1v}$) and asymmetric particles with no mirror planes (trivial group $C_1$), as revealed by birefringent textures and defocusing of different parts of the particle. An interesting interplay between symmetries of distortions in $\mathbf{n}(\mathbf{r})$ and $\chi(\mathbf{r})$ can be noticed when comparing like-shaped particles introduced into nematic and cholesteric LCs, respectively, as we discuss in the Supplemental Material [38]. Furthermore, since four types of pure and many hybrid elastic dipoles in nematic LCs can be identified using a tensorial description based on different symmetries of elastic distortions [11], we extend this description to CLCs and then further classify elastic dipoles obtained in our experiments in the Supplemental Material [38,39]. In both nematic and cholesteric LCs, LCE colloids rotate to minimize elastic energy of the surrounding $\mathbf{n}(\mathbf{r})$ or $\chi(\mathbf{r})$ in response to changes of particle shape (video 1 in [38]) and even the simplest nonreciprocal dynamic morphing of the shape results in a directional locomotion (video 2 in [38]).

To conclude, we have introduced a new class of active LCE particles with light-controlled rigidly connected fragments of cylinders in fluid LCs that can be described by employing analogies with charge distributions in electrostatics. These elastomeric colloids will allow one to explore light-driven rotational and translational motion in low-viscosity nematic and cholesteric LCs of varying pitch and will bridge the fields of LC solids and active colloids. Furthermore, our double-LC colloidal system may allow one to explore the role of geometric shape in determining the complex dipolar structures, light-controlled self-assembly, reconfigurable knotting of defects, and formations of LC gels [37,39–41]. Potential applications include laser-guided assembly of reconfigurable materials for photonic and electro-optic applications and light-driven micromachines.

We acknowledge the support of NSF Grant No. DMR-0847782, Grant No. P200A030179 of the U.S. Department of Education, and ANR Grant No. ANR-2010-INTB-904-01, as well as technical assistance of P. Ackerman, Z. Qi, and R. Trivedi.

**Figures**

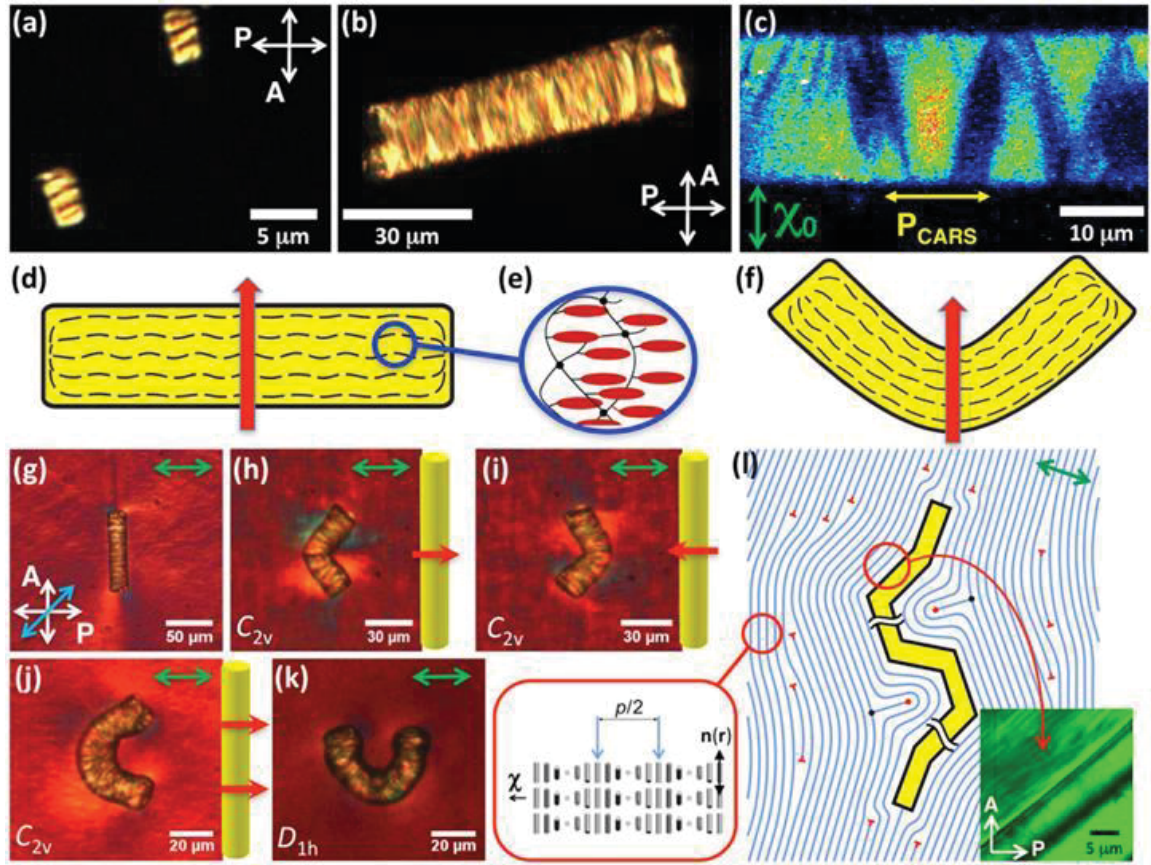

FIG. 1. LCE particles in isotropic and LC hosts. (a),(b) POM images of LCE particles in water. (c) CARS-PM image of an LCE cylinder in a lyotropic LC obtained for a linear laser polarization marked by a yellow double filled arrow. (d), (e), (f) $\mathbf{n}(\mathbf{r})$ in LCE particles (d) before and (f) after the scanning along direction marked by red arrows. (g) LCE cylinder in a CLC. (h),(i) Bent cylinders obtained by laser scanning along directions shown by red arrows in the insets. (j),(k) LCE particles shaped into letters (j) C and (k) U by means of laser scanning along red arrows shown in (j). Crossed polarizer $P$ and analyzer $A$ are marked by white crossed double arrows while the "slow" axis of a 530 nm retardation plate is shown by a blue tilted double arrow in (g) and correlates with the darker blue color in the texture, visualizing regions where CLC layers are roughly along the blue tilted double arrow; green double opened arrows mark $\chi_0$. (l) Schematic of an optically shaped LCE colloid distorting CLC and introducing edge dislocations (red nails); cylinders in the inset represent $\mathbf{n}(\mathbf{r})$. POM image in the inset of (l) shows that CLC layers are parallel to the particle's surface.

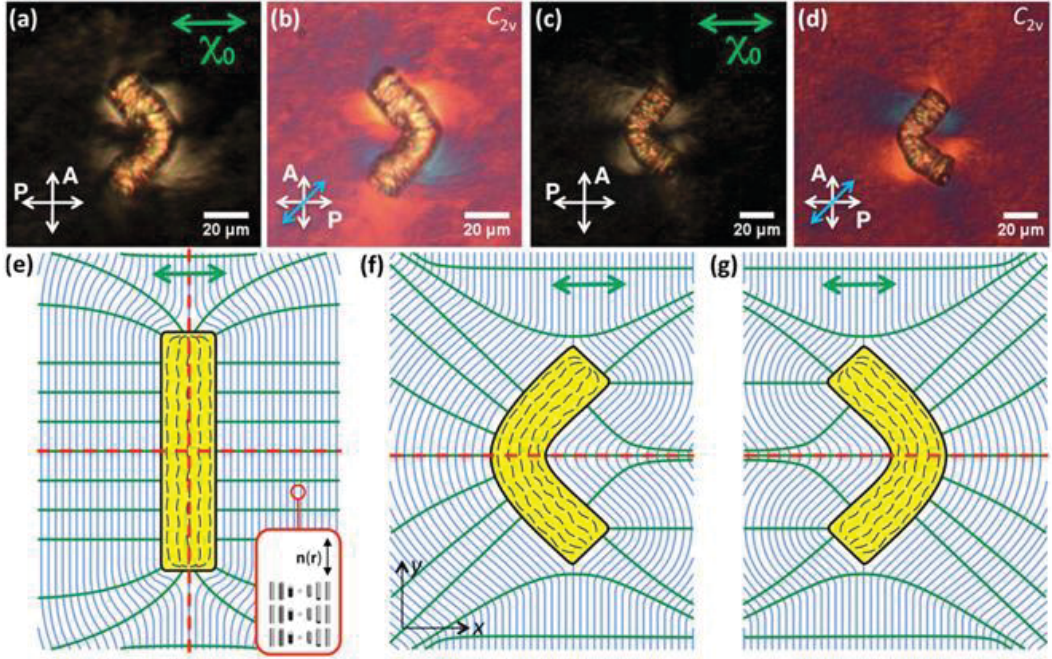

FIG. 2. Control of $\chi(\mathbf{r})$- symmetry and layered structures in a CLC via optical control of LCE colloidal shapes. (a)–(d) POM images of LCE particles obtained (a),(c) without and (b),(d) with a 530 nm retardation plate. (e) $\mathbf{n}(\mathbf{r})$ within the cylindrical LCE particle (black dashed lines), and the layer structures (blue thin lines) and $\chi(\mathbf{r})$ (green thick lines) around it have mirror symmetry planes marked by dashed red lines as well as coinciding with the plane of the schematic. Cylinders in the inset represent $\mathbf{n}(\mathbf{r})$ twisting by $\pi$ within each cholesteric layer. (f),(g) Layer structures and $\chi(\mathbf{r})$ induced via bending of LCE particles.

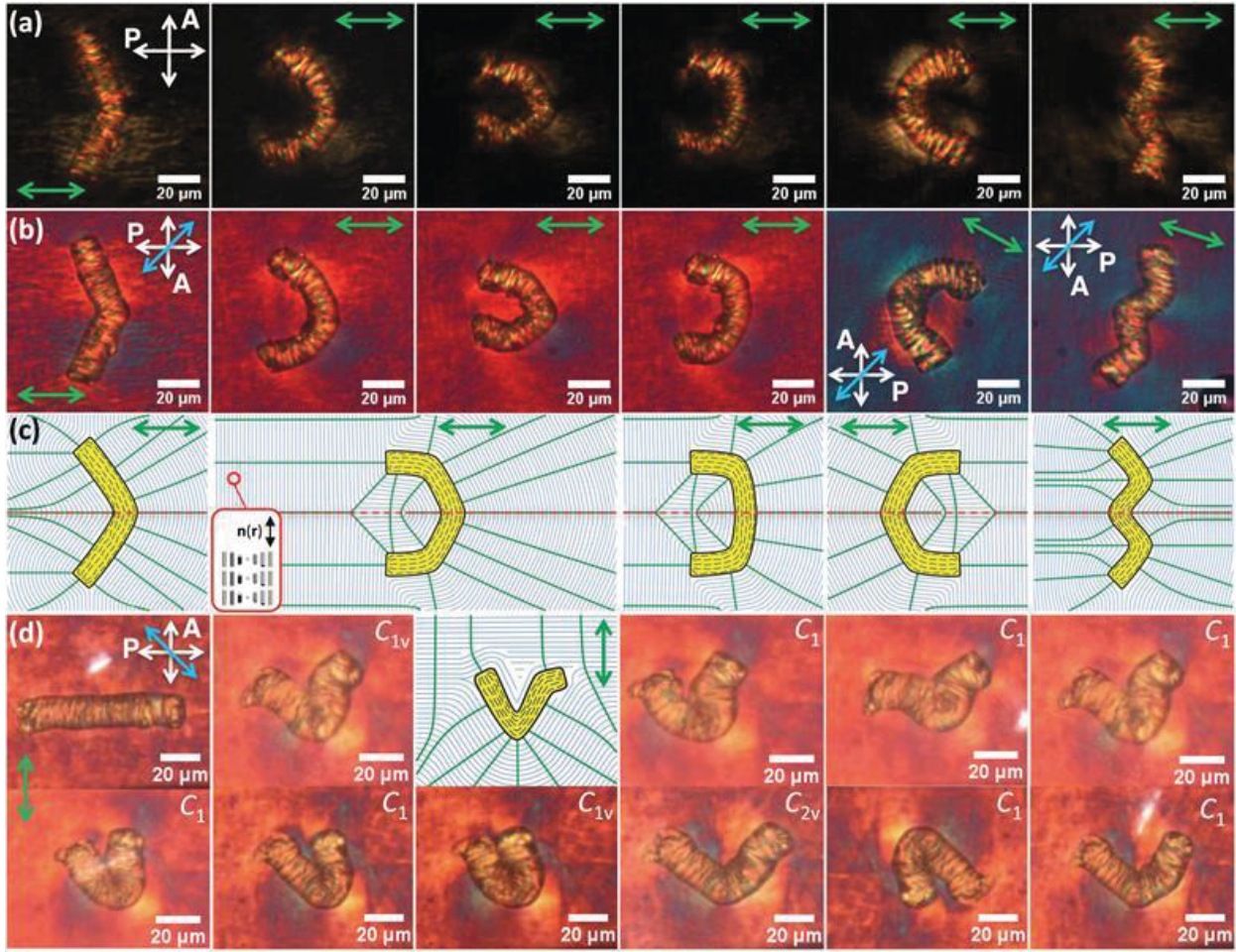

FIG. 3. Low-symmetry χ(**r**) structures induced by continuous optical morphing of LCE colloids. (a),(b) POM images of LCE particles and χ(**r**) in the surrounding LC host obtained (a) without and (b) with a 530 nm retardation plate. (c) Schematics of cholesteric layers and χ(**r**) around LCE particles corresponding to POM images shown in (a),(b). (d) POM images obtained with a retardation plate showing examples of shapes and χ(**r**) configurations obtained by morphing LCE particles in a CLC. The last two images on the right side of (b) were taken when the sample was rotated with respect to crossed polarizers.

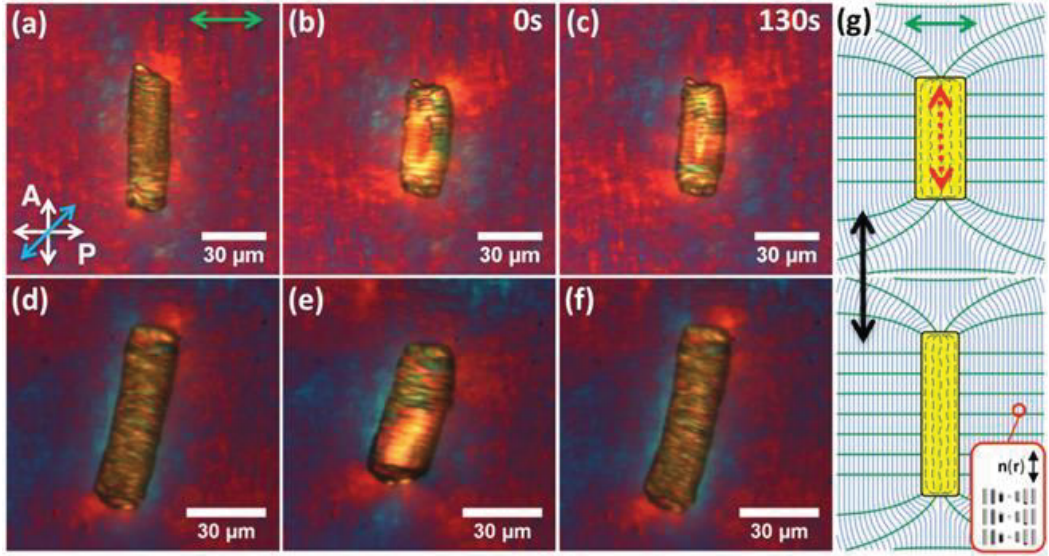

FIG. 4. Reversible and irreversible control of aspect ratios of LCE cylinders in a CLC. (a) LCE cylinder in a sheared CLC. (b),(c) Length shrinking and diameter increase induced by bidirectional laser scanning along the cylinder axis that retains $\chi(\mathbf{r})$ symmetry and persists with time (marked in top right corners of POM images) after scanning is discontinued abruptly. (d) LCE particle in a non-sheared CLC. (e), (f) Reversible shape modification induced by similar bidirectional scanning but terminated via a continuous decrease of laser power within ~2 min and accompanied by reversible change of $\chi(\mathbf{r})$ in the CLC. (g) A schematic showing changes in $\chi(\mathbf{r})$ and layer structures that follow laser-induced modification of the particle shape; the dashed red double arrow depicts bidirectional scanning of the beam. The modified shape of particle is long-term stable if scanning is discontinued via turning laser off, but relaxes back to the original one when intensity of the beam decreases gradually.

# Supplemental Material

# Active shape-morphing elastomeric colloids in short-pitch cholesteric liquid crystals

J. S. Evans, Y. Sun, B. Senyuk, P. Keller, V. M. Pergamenshchik,

T. Lee, and I. I. Smalyukh

## I. OUTLINE OF THE SUPPLEMENT

Here we discuss a parallel between the symmetry-based identification of different dipolar colloidal particles in a nematic liquid crystal (NLC) and that in a short-pitch cholesteric liquid crystal (CLC). The idea is to find a connection between weak distortions of the nematic director and weak distortions of the field of normals to the cholesteric layers. First we briefly present the main expressions of the elastic multipole expansion, the results of the symmetry-based classification of dipolar colloidal particles [1], and describe the related symmetry elements in a NLC. Then we sketch the way to establish the elastic multipole expansion in a CLC. As a result, we find a similarity between the tensorial structures of elastic dipoles in both LCs and argue that colloidal particles in a CLC can be described in a way similar to the tensorial description in colloidal nematostatics. Finally, we illustrate the similarities and differences between the two descriptions by sketching the distortions around the particles which were experimentally realized in the paper.

## II. THE SYMMETRY-BASED SIMILARITY BETWEEN DIPOLAR DISTORTIONS OF THE n-DIRECTOR IN A NLC AND THE $\chi$-DIRECTOR IN A SHORT-PITCH CLC

### A. The director expansion in a NLC

The full theory of a large-distance distortion-mediated interaction of colloidal particles in a CLC has not been developed so far. At the same time, the symmetry-based classification of different dipolar particles and their interactions have been derived for a nematic LC in Ref. [1]. Here we show that it is possible to establish a similarity between the symmetries associated with different distortion sources in a NLC and in a CLC, in spite the great difference in their ground states. To this end we first briefly present some instructive facts from colloidal nematostatics and then sketch how the multipole expansion can be developed for a CLC. As a result, we find certain similarity between the two theories. Finally, we illustrate the symmetry associated with dipolar particles in a CLC and NLC by sketching different director fields.

The director ground state in a NLC is homogeneous. Let a 3D director field $\mathbf{n}(\mathbf{r})$, uniform and parallel to the $z$-axis at infinity, $\mathbf{n}_0 = (0,0,1)$, be weekly distorted by a particle. Enclose this particle by a spherical surface $S$ of radius $a$, and let $\nu(\mathbf{s})$ be the vector of unit normal to $S$ at its point $\mathbf{s}$ (directed inward the sphere). The perturbation $\delta \mathbf{n}$ of $\mathbf{n}_0$ is small, $|\delta \mathbf{n}| \ll 1$, and transverse, $\delta \mathbf{n} = (\delta n_x, \delta n_y, 0)$. In the one constant approximation the free energy of these distortions outside the sphere is

$$F_{NLC} = \frac{K}{2} \int_{V_{out}} (\nabla \delta n_t \cdot \nabla \delta n_t) d^3 V \qquad (1)$$

(throughout this text, a summation over repeated indices is implied). The field $\delta n_t(\mathbf{r})$, $t = x, y$, satisfies Laplace's equation. Its solution in the region with the inner boundary $S$, that vanishes at infinity, is uniquely determined by the director distribution $\delta n_t$ on $S$. This solution can be found and expanded in the inverse powers of the distance $R$ from the origin which can be reduced to the following form [2, 3]:

$$n_t(\mathbf{r}) = \frac{q_t}{R} + 3\frac{(\mathbf{d}_t \cdot \mathbf{R})}{R^3} + ..., \qquad (2)$$

where

$$q_t = \frac{1}{4\pi a} \int_S n_t d^2 s, \qquad (3)$$

$$d_{t,\alpha} = -\frac{1}{4\pi} \int_S n_t \nu_\alpha d^2 s. \qquad (4)$$

The sum (2) is the elastic multipole expansion and Eqs. (2)-(4) determine the elastic multipoles. The form of (2)-(4) prompts following interpretation. $q_t$ is the $t$-th component of the elastic charge, the object $\mathbf{d}_t$ is the $t$-th dipole moments determined (in quite the standard way) by the surface "charge density" $\propto n_t$ on the sphere. The two separate sources $n_x$ and $n_y$ determine not only the $x$ and $y$ director components outside the particle, but also two components, the tensor dyad ([5], pp.157 and 284), for each multipole moment, i.e., $q_x$ and $q_y$, $\mathbf{d}_x$ and $\mathbf{d}_y$, .... The dyadic form (3) and (4) is pertinent to our peculiar reference frame with the $z$-axis along the undistorted director. In a general reference frame, the elastic charge is characterized by a 3D vector $q_\alpha$ and the elastic dipole is characterized by a second rank 3D tensor $d_{\alpha\beta}$, where all indices run over $1,2,3$. However, as Eqs. (3),(4) determine the general form of an elastic multipole up to an arbitrary rotation in 3D space, it is most natural to consider the multipole structure in the reference frame with $n_z = 1$, where $q_3 = d_{3,\beta} = 0$ and the parameters related to any particular rotation of the symmetry axis $\mathbf{n}_\infty$ are excluded. Here we use only this reference frame. Then $d_{t,\alpha}$ is the $2 \times 3$ matrix $||d_{t,\alpha}||$ whose first and second rows coincide, respectively, with the components of $\mathbf{d}_x$ and $\mathbf{d}_y$, i.e.,

$$||d_{t,\alpha}|| = \begin{pmatrix} d_{x,x} & d_{x,y} & d_{3,x} \\ d_{y,x} & d_{y,y} & d_{3,y} \end{pmatrix} = \begin{pmatrix} \mathbf{d}_x \\ \mathbf{d}_y \end{pmatrix}. \tag{5}$$

This object can be decomposed into the second rank $2 \times 2$ tensor $\mathbf{D}$ and the 2D vector $\mathbf{d}_3$: $(\mathbf{d}_x, \mathbf{d}_y) = (\mathbf{D} \oplus \mathbf{d}_3)$ where

$$\mathbf{D} = ||d_{t,t'}|| = \begin{pmatrix} d_{x,x} & d_{x,y} \\ d_{y,x} & d_{y,y} \end{pmatrix}, \tag{6}$$

$$\mathbf{d}_3 = (d_{3,x}, d_{3,y}). \tag{7}$$

This separation is justified because under all of the transformations of interest, described below, $\mathbf{D}$ and $\mathbf{d}_3$ are transformed independently, and $||d_{t,\alpha}||$ (5) is zero if and only if $\mathbf{D} = 0$ and $\mathbf{d}_3 = 0$. As a rotation about the homogeneous director $\mathbf{n}_\infty$ does not alter the free energy of a particle, all the elastic multipoles should be determined up to an arbitrary azimuthal angle. Choosing this angle, $\mathbf{D}$ can be reduced to the special form [1, 4]

$$\mathbf{D} = d\mathbf{d} + \Delta\boldsymbol{\Delta} + C\mathbf{C}, \tag{8}$$

where

$$\mathbf{d} = \begin{pmatrix} 1 & 0 \\ 0 & 1 \end{pmatrix},$$

$$\boldsymbol{\Delta} = \begin{pmatrix} 1 & 0 \\ 0 & -1 \end{pmatrix}, \tag{9}$$

$$\mathbf{C} = \begin{pmatrix} 0 & 1 \\ -1 & 0 \end{pmatrix}.$$

The quantity $\mathbf{d}_3$ (7) transforms as a 2D vector: $(d_{3,x}, d_{3,y}) \to (\gamma, \gamma')$. Thus, in this eigenframe the dipolar dyad, which fully determines the distortions of the source (as well as its interaction with other multipoles [2, 3]) takes the form

$$\begin{align} \mathbf{d}_{0x} &= (d + \Delta,\ C,\ \gamma), \\ \mathbf{d}_{0y} &= (-C,\ d - \Delta,\ \gamma'). \end{align} \tag{10}$$

We see that, in general, an elastic dipole can be described by three coefficients $d$, $\Delta$, $C$, and a two component vector $(\gamma, \gamma')$. We call the coefficients $d$, $\Delta$, and $C$ isotropic, anisotropic, and chiral dipole strength, respectively; the vector $\mathbf{d}_3$ is a longitudinal dipole as it comprises the dyad's components along $\mathbf{n}_\infty$. Below we discuss a connection between $d$, $\Delta$, $C$, $\gamma$, $\gamma'$ and symmetry of director distortions induced by a source particle. But first we show that the same tensorial structure appears in relation with particle-induced distortions in a CLC.

### B. The $\chi$-director expansion in a CLC

Consider a short-pitch CLC within the framework of the Lubensky-de Gennes approach [5]. The ground state with the flat cholesteric layers normal to the $z$-axis is weakly distorted by a particle. The field of unit normals to the layers now has the form $\boldsymbol{\chi} = (\delta\chi_x, \delta\chi_y, 1)$ where $\delta\boldsymbol{\chi}$ is a small transverse perturbation of the ground state $\boldsymbol{\chi}_0 = (0, 0, 1)$. The layers are displaced by $u$, which is connected to the vector $\delta\boldsymbol{\chi}$ via $\delta\boldsymbol{\chi} = -\nabla_\parallel u$ where $\nabla_\parallel$ is the gradient in the $xy$ plane. The free energy of the model can be written in the form

$$F_{CLC} = \frac{1}{2} \int d^3\mathbf{x} [B(\partial_z u)^2 + K_1 (\nabla \cdot \delta\boldsymbol{\chi})^2], \tag{11}$$

where $B$ is the compressibility modulus of the cholesteric layers, $K_1$ is the effective splay constant of the $\boldsymbol{\chi}$ director, and $\partial_z = \partial/\partial z$. Minimization of this functional in different situations has been considered in a number of publications (e.g., see [6–9]). It was shown in Refs. [6, 7] that the displacement $u$ induced by a particle can be imposed by means of introducing the field $\psi(\mathbf{x})$ (Lagrange multiplier). This field resides in the area $\Sigma$ of the particle and is coupled to the layer compression $\partial_z u$. As a result, the displacement can be expressed as the following integral over $\Sigma$ [7]:

$$u(\mathbf{x}) = \int_\Sigma d^3\mathbf{x}' G(\mathbf{x} - \mathbf{x}') \psi(\mathbf{x}'), \tag{12}$$

where $G$ is some kernel setting the distance dependence of $u$ from a point-like source [counterpart of $1/R$ in the expansion (2)]. A multipole expansion for the field of the surface normals $\delta\boldsymbol{\chi} = -\nabla_\| u$ can be obtained into two steps. First, one applies the operator $\nabla_\|$ to both sides of formula (12) and then by-part integrates the right hand side. Second, since far from the particle $|\mathbf{x}|$ is large comparing to the size of $\Sigma$, the integrand of (12) is expanded in powers of $\mathbf{x}'$ which gives a multipole expansion for $\nabla_\| u$. As a result, the transverse components $\delta\chi_t$ can be written as follows:

$$\delta\chi_t(\mathbf{x}) \propto \widetilde{q}_t G(\mathbf{x}) + \widetilde{d}_{t,\alpha} \partial_\alpha G(\mathbf{x}) + ... \tag{13}$$

Here $\partial_\alpha = \partial/\partial x_\alpha$, $t = x$ or $y$, $\alpha$ and $\beta$ run over $x, y$, and $z$, and the tensorial coefficients have the form

$$\widetilde{q}_t = -\int_\Sigma \partial_t \psi(\mathbf{x}') d^3\mathbf{x}', \tag{14}$$

$$\widetilde{d}_{t,\alpha} = \int_\Sigma \partial_t \psi(\mathbf{x}') x'_\alpha d^3\mathbf{x}'. \tag{15}$$

The applicability of the approach based on the multipole expansion in a short pitch CLC has certain limitations. First, it is generally known that any multipole expansion is a practically useful tool only when the distance between sources is much larger than their sizes. In particular, the symmetry-based consideration above as well as the classification of Ref.[1] assume weak deformations that take place at suffieciently large distances from the source. At short distances, however, the deformations are not generally weak and cannot be described by any single multipole field. For instance, stable static colloid assemblies with short interparticle distances, which are usually accompanied by complicated defect lines, can be described only numerically (e.g., see recent Ref.[10]). The second limitation follows from that of the Lubensky-de Gennes approach operating with the field of normals. The description in terms of the field of normals to the cholesteric layers implies that the layer thickness $p/2$ (distance over which director rotates by $\pi$) is the shortest length scale of the problem and is very small in comparison with the characteristic distortion length. Therefore, cholloidal particles that can be treated in terms of the Lubensky-de Gennes approach have to be much larger than the cholesteric pitch. If a particle size is of the order or smaller than the pitch $p$, then the very idea of "cholesteric layers" is meaningless as the distortions change considerably within the distance smaller than or comparable with the cholesteric pitch (i.e., within single "layer"). In this case, the description in terms of the field of normals, used in this study, cannot be applied. In our case, however, the assumptions of the Lubensky-de Gennes approach are satisfied: all of the particles we studied in a CLC are considerably larger than the cholesteric pitch which justifies our resorting to the field of normals.

### C. The similarity

We see that there is a similarity between the **n** expansion (2) and the $\boldsymbol{\chi}$ expansion (13): the coefficients $q_t$ and $\widetilde{q}_t$, $d_{t,\alpha}$ and $\widetilde{d}_{t,\alpha}$, i.e., the elastic multipoles in a NLC and the expansion coefficients in a CLC have similar tensorial structures. In particular, $\widetilde{d}_{t,\alpha}$ (15) has the same structure as $d_{t,\alpha}$ (4), hence $\widetilde{d}_{t,\alpha}$ can be reduced to the form

$$||\widetilde{d}_{t,\alpha}|| = \begin{pmatrix} \widetilde{\mathbf{d}}_x \\ \widetilde{\mathbf{d}}_y \end{pmatrix} = (\mathbf{D}_\chi \oplus \mathbf{d}_{3\chi}), \tag{16}$$

where

$$\begin{aligned} \mathbf{D}_\chi &= d_\chi \mathbf{d} + \Delta_\chi \boldsymbol{\Delta} + C_\chi \mathbf{C}, \\ \mathbf{d}_{3\chi} &= (\gamma_\chi, \gamma'_\chi), \end{aligned} \tag{17}$$

with the same tensors $\mathbf{d}$, $\boldsymbol{\Delta}$, and $\mathbf{C}$ given in (9). Thus, a distortion of the $\boldsymbol{\chi}$ director induced by a dipolar particle in a short-pitch CLC can be specified by coefficients $d_\chi, \Delta_\chi, C_\chi$, and a vector $(\gamma_\chi, \gamma'_\chi)$ which have the same symmetry-based meaning as their counterparts in a NLC. It should be stressed however that in a NLC these coefficients not only describe the particle-induced distortion, but also fully determine the dipole-dipole interaction. If the dipole-dipole interaction in a CLC is fully determined by these coefficients as well is presently open to question.

### D. The symmetry groups of the constituent dipolar tensors and correspondent director distortions

The undistorted homogeneous directors $\mathbf{n}_0$ and $\boldsymbol{\chi}_0$ have the same symmetry group $D_{\infty h}$ which comprises any rotations about the $z$-axis, the reflection in any vertical mirror plane (VMP) passing through the $z$-axis, the reflection in a horizontal mirror plane (HMP), and the rotations by an angle $\pi$ about any horizontal axis. The director field induced by any particle cannot have a higher symmetry: the maximum symmetry $D_{\infty h}$ remains intact only far from the distortion source. Therefore, considering possible symmetries of a particle-induced director field, one has to restrict possible transformations to those of $D_{\infty h}$.

A symmetry element of a director field is a transformation which does not alter this field. Similarly, a symmetry element of a tensor is a transformation which does not alter the tensor. As the symmetry of $\widetilde{d}_{t,\alpha}$ is determined by that of $\boldsymbol{\chi}$, both symmetries are connected (see Ref.[1] for details). Each of the three tensors in (9) and the vector $\mathbf{d}_{3\chi}$ has its particular symmetry elements; a particular tensor in the sum (17) appears with a nonzero coefficient only if the symmetry elements of the director field are among those of this particular tensor. In particular, the tensor has a nonzero coefficient if the director symmetry group is a subgroup of its maximum symmetry group.

In what follows we use the notations and description of the point groups from the book [11] (the identical transformation present in any group will not be mentioned). The maximum symmetry group $G_d$ of the tensor $\mathbf{d}$ is $C_{\infty v}$ [rotation about the $z$-axis by any angle and a VMP passing thorough the $z$-axis and any horizontal axis]. The symmetry $C_{\infty v}$ corresponds to the full isotropy of all the directions normal to this $z$-axis, hence $d$ is called isotropic strength of an elastic dipole. The coefficient $d$ is nonzero if the director is invariant with respect to subgroups of $C_{\infty v}$. The important subgroups of $G_d$ are the groups $C_n$ [$n$ $z$-rotations by an angle $2\pi/n$] and $C_{nv}$ [$n$ $z$-rotations by angle $2\pi/n$ and $n$ VMPs] of any order $n$.

The maximum symmetry group $G_\Delta$ of the tensor $\boldsymbol{\Delta}$ is $D_{2d}$. The eight symmetry operations of the tensor $\boldsymbol{\Delta}$ are: those of the group $C_{2v}$ (two $z$-rotations through $\pi$ and two reflections in the two VMPs), $\pi$ rotation about the horizontal axis $\zeta_x$ lying in the midway between the two VMPs, $\pi$ rotation about the horizontal axis $\zeta_y$ perpendicular to $\zeta_x$, and the two rotary reflection transformations from the group $S_4$ [reflection in a HMP followed by the rotation through an angle $2\pi/4$ about the vertical $z$-axis]. The symmetry $C_2$ is related to the two preferred directions which violate the high rotational symmetry of the isotropic $d$ tensor (and that of the chiral $C$ tensor), and $\Delta$ is called anisotropic strength of an elastic dipole. The important subgroups of $G_\Delta$ are $S_4$ and $V$ ($V \equiv D_2$ has three mutually perpendicular second order rotary axes, the two horizontal $\zeta$ axes and one vertical).

The symmetry group of the tensor $\mathbf{C}$ is $D_n$ which consists of $n$ $z$-rotations from $C_n$ and $n$ rotations through an angle $\pi$ about $n$ horizontal axes. The maximum symmetry group $G_C$ of the tensor **C** is $D_\infty$. The absence of reflection planes and rotary reflection axes, but, at the same, the invariance with respect to rotations is attributed to the chiral symmetry and helicoidal structures, and $C$ is called chiral strength of an elastic dipole. The groups $C_n$ and $D_n$ with any $n$ as well as the group of a $\pi$-rotation $U_2$ about any horizontal axis are important subgroups of $G_C$.

The maximum symmetry group $G_{d_3}$ of the vector $\mathbf{d}_3$ is $D_{1h}$ (a HMP, the second order horizontal rotary axis $x'$, and a VMP passing through this axis). As $\mathbf{d}_3$ is the only component of the dipole diad (5) or (16) which is along the unperturbed director, we call it $z$ or $\gamma$ dipole. The sub-groups of $G_{d_3}$ are $U_2$ (the $\pi$ rotation about a horizontal axis $x'$), $C_{1h} \equiv C_s$ (HMP reflection), and $C_{1v}$ (reflection in the VMP passing through the principal axis $x'$).

## III. SKETCHES OF THE ELASTIC DIPOLES REALIZED IN A CLC AND THEIR ROLE IN A NLC

The different dipolar particles experimentally realized in the paper are shown in Figs. 1h-k, 2a-d and 3. Let us group these particles into different types according to the symmetry of the $\chi$ distortions around them. Let the undistorted field $\chi_0$ of normals to the cholesteric layers be along the $z$-axis. The V and U shaped particles in Figs. 1h-j, 2a-d, and 3a-c (and even the zig-zag V-shaped particles in the right columns of Fig. 3a-c) have the following symmetries: $z$-rotations by an angle $\pi$, a VMP lying in the plane of the particle's "V" and passing through the $z$-axis. These symmetry elements are those of the group $C_{2v}$ which is a sub-group of both maximum groups $G_d$ and $G_\Delta$, hence they can be described by two coefficients $d_\chi$ and $\Delta_\chi$. This is a biaxial $\chi$-dipole or $(d_\chi, \Delta_\chi)$ dipole. The symmetry of the $\chi$ field induced by this $\chi$-dipole is sketched in the $xz$ and $xy$ planes in Fig. S1a. One VMP is passing through the $x$-axis, and the other one through the $y$-axis. For comparison, Fig. S1b presents the same information for the **n** field induced by a biaxial $(d, \Delta)$ **n**-dipole in a NLC. The symmetries of the $\chi$ field and **n** field are obviously the same which justifies our idea of the similarity in the dipolar types. If instead of the ellipse-like biaxiality, the $xy$ field projection has the circular symmetry, then the total symmetry group is $C_{\infty v}$, $\Delta = 0$, and one has a pure isotropic dipole described by a single coefficient $d$.

In principle, the colloidal particles obtained experimentally in a CLC can be used in a NLC as well. The surface anchoring of the particle remains the same in both LCs, but the particle type changes as in the former case it is determined by the symmetry of the $\boldsymbol{\chi}$ director whereas in the later case it is determined by the **n** director. For instance, if the $\boldsymbol{\chi}$ director is normal to the particle surface, then the **n** director is tangential to it and vise versa. In the case of a V shaped particle, this implies that the angles the apex direction makes with the **n** and $\boldsymbol{\chi}$ directors always differ by $\pi/2$: if the apex of the V points in the direction normal to $\boldsymbol{\chi}$ in a CLC, then it points along **n** in a NLC. This is illustrated in Fig. S1c for the nematic obtained from the CLC of Fig. S1a by unwinding its spiral. It is seen that the **n** director field in Fig. S1c, which is induced in a NLC by the particle shown in Fig. S1a, has the symmetry different from that of the $\boldsymbol{\chi}$ field in Fig. S1a. This symmetry which will be described below is that of the vector $\mathbf{d}_3$ (7), and the **n**-dipole in Fig. S1c is a pure $\gamma$ dipole.

Now consider the particle shown in Fig. 1k. The $xz$ (vertical) projection of the $\boldsymbol{\chi}$ field induced by it is sketched in Fig. S2a. The symmetry elements of the $\boldsymbol{\chi}$ field shown in Fig. 2a are a VMP lying in the particle's plane, a HMP $z = 0$ normal to this VMP and passing through the apex of the "U", and a $\pi$ rotation about the horizontal $x$-axis. These are the elements of the group $D_{1h}$ which is the maximum symmetry group of the vector $\mathbf{d}_3$. Thus, the particles in Fig. 1k and Fig. S2a are pure $\gamma_\chi$ dipole. For comparison, Fig. S2b presents the same information for the **n** field induced by a $\gamma$ dipole in a NLC. It is clear that this is the same particle as in Fig. S1c, so that the $xy$ projection in Fig. S1c is pertinent to Fig. 2a,b. The particle of Figs. 1k and S2a, which is a $\gamma_\chi$ dipole in a CLC, would be a $(d, \Delta)$ **n**-dipole in a NLC. This is illustrated in Fig. S2c for a NLC which could be obtained by unwinding the spiral of the CLC shown in Fig. S2a: the symmetry of the **n** director is obviously the same as that in the Fig. S1b.

We see that there is certain duality: a $\gamma_\chi$ dipole in a CLC would be a $(d, \Delta)$ dipole in a NLC (Fig. S2a and c), and vise versa, a $(d_\chi, \Delta_\chi)$ dipole in a CLC would be a pure $\gamma$ dipole in a NLC.

Finally, the above consideration show that bend-cylinder particle-induced dipolar $\boldsymbol{\chi}$ distortions can be characterized by coefficients $d_\chi, \Delta_\chi, C_\chi$, and $\gamma_\chi$ analogous to their nematic counterparts. A particle which is quadrupolar in a CLC remains quadrupolar in a NLC, compare Figs. 2e and S3a. However, the $\boldsymbol{\chi}$ distortions in a CLC and the **n** distortions in a NLC induced by the same dipolar particle have different symmetries implying that the same dipolar particle is of different types in these two phases. For instance, bend-cylinder particles with the cusp along $\boldsymbol{\chi}_0$ shown in Figs. 1h-j, 2a-d, f, g and 3a-c are mixed $(d_\chi, \Delta_\chi)$ dipoles, although they would be $\gamma$ dipoles in a NLC as their cusps would be normal to $\mathbf{n}_0$, Figs. S3b-d. A V-shaped particle with a cusp normal to $\boldsymbol{\chi}_0$ shown in Fig. 1k is a $\gamma_\chi$ dipole, while it would be a $(d, \Delta)$ dipole in a NLC. The asymmetric dipolar particles in Fig. 3d can be characterized as mixed type $\boldsymbol{\chi}$ dipoles with nonzero $d_\chi, \Delta_\chi, \gamma_x$, and possibly a small chirality $C_\chi$, but with the dominating $(d_\chi, \Delta_\chi)$ component. Note that changing the aspect ratio does not change the quadrupolar character of a particle both in CLC and in NLC, Fig. S4.

## IV. FIGURE CAPTIONS

**Figure S1.** Sketches of the distortions induced by yellow particle in a NLC and CLC. The three upper sketches show the distortions in the vertical plane passing through particle's apex ("vertical" direction is that of the $z$-axis which is along the unperturbed homogeneous director lines): (a) Distortions of the $\boldsymbol{\chi}$ director (green lines) and of the cholesteric planes (blue lines) induced by a flatten-cone-shaped particle with planar anchoring all over its surface. The inset shows a helical alignment of the $\mathbf{n}$ director in the blue layers. The lower sketch is the horizontal projection, it shows small transverse distortions $\delta\boldsymbol{\chi}$ of the vertical lines (i.e., of $\boldsymbol{\chi}_0$). The symmetry group of the green $\delta\boldsymbol{\chi}$ arrows is $C_{2v}$ which contains a VMP passing through the $x$-axis, a VMP passing through the $y$-axis, and two $\pi$ rotation about the $z$-axis. By analogy with the colloidal nematostatics [1], this particle in a CLC can be identified as a $(d_\chi, \Delta_\chi)$ doublet (see caption below). (b) Distortions of the $\mathbf{n}$ director (blue lines) which have exactly the same symmetry $C_{2v}$ as the distortions of the $\boldsymbol{\chi}$ director in Fig. 1Sa. The distortions are induced by a flatten-cone shaped particle with a planar anchoring on the lateral surface and normal anchoring on its base. The lower sketch is the horizontal projection, it shows small transverse distortions $\delta\mathbf{n}$ of the vertical lines (i.e., of $\mathbf{n}_0$). According to the classification of nematic elastic dipoles, this is a $(d, \Delta)$ doublet [1]. By analogy, the particle in a CLC, shown in Fig. S1a, can be identified as a $(d_\chi, \Delta_\chi)$ doublet. If the picture in the lower sketch is circular, then $\Delta = 0$. The well-known particular case of a pure isotropic $d$ dipole is the topological dipole shown in the inset. (c) Distortions of the $\mathbf{n}$ director induced in a NLC by the particle of Fig. S1a. In particular, such a NLC can be obtained by unwinding the cholesteric spiral of Fig. S1a. The lower sketch shows the $yz$

projection. The symmetry group of the **n** director is $D_{1h}$. It has a VMP (the $xz$ plane), a HMP (the $xy$ plane), and a $\pi$ rotation about the $x$-axis. This nematic elastic dipole is identified as a $\gamma$ dipole [1].

**Figure S2.** Sketches of the distortions induced by yellow particles in a NLC and CLC. The three sketches show the distortions in the vertical plane passing through particle's apex ("vertical" direction is that of the $z$-axis which is along the unperturbed homogeneous director lines): (a) Distortions of the $\boldsymbol{\chi}$ director (green lines) and of the cholesteric planes (blue lines) induced by a V shaped particle with planar anchoring all over its surface. The $xz$ projection (not shown) is similar to that in Fig. S1c. The symmetry group of the green $\boldsymbol{\chi}$ lines is $D_{1h}$ (see caption to Fig. S1c). A dipolar particle with the same symmetry relative to the **n** director, Fig. S1c, is called $\gamma$ dipole [1]. By analogy with the nematostatics, the $\boldsymbol{\chi}$ dipole can be identified as a $\gamma_\chi$ dipole. (b) Distortions of the **n** director (blue lines) which have exactly the same symmetry $D_{1h}$ as the distortions of the $\boldsymbol{\chi}$ director in Fig. 2Sa. The distortions are induced by a flatten-cone-shaped particle with a planar anchoring all over its surface. This is a $\gamma$ dipole [1]. (c) Distortions of the **n** director induced in a NLC by the particle of Fig. S2a. In particular, such a NLC can be obtained by unwinding the cholesteric twist of Fig. S2a. The symmetry group of the **n** director is $S_{2v}$ (see caption to Fig. S1a), and the particle is a nematic $(d, \Delta)$ doublet.

**Figure S3.** Schematic diagrams showing distortions induced by bend-cylinder particles in a NLC. (a) Quadrupolar colloidal particle. Quadrupolar particles are quadrupolar in either phase. (b-d) Bend-cylinder particles with the cusp normal to $\mathbf{n}_0$ forming $\gamma$ dipoles. Note that similar bend-cylinder particles but with the cusp along $\boldsymbol{\chi}_0$ in a CLC shown in Figs. 1h-j, 2a-d, f, g and 3a-c are mixed $(d_\chi, \Delta_\chi)$ dipoles. The **n** director is shown by blue lines.

**Figure S4.** Quadrupolar particles with different laser-controlled aspect ratios in a NLC. The **n** director is shown by blue lines.

---

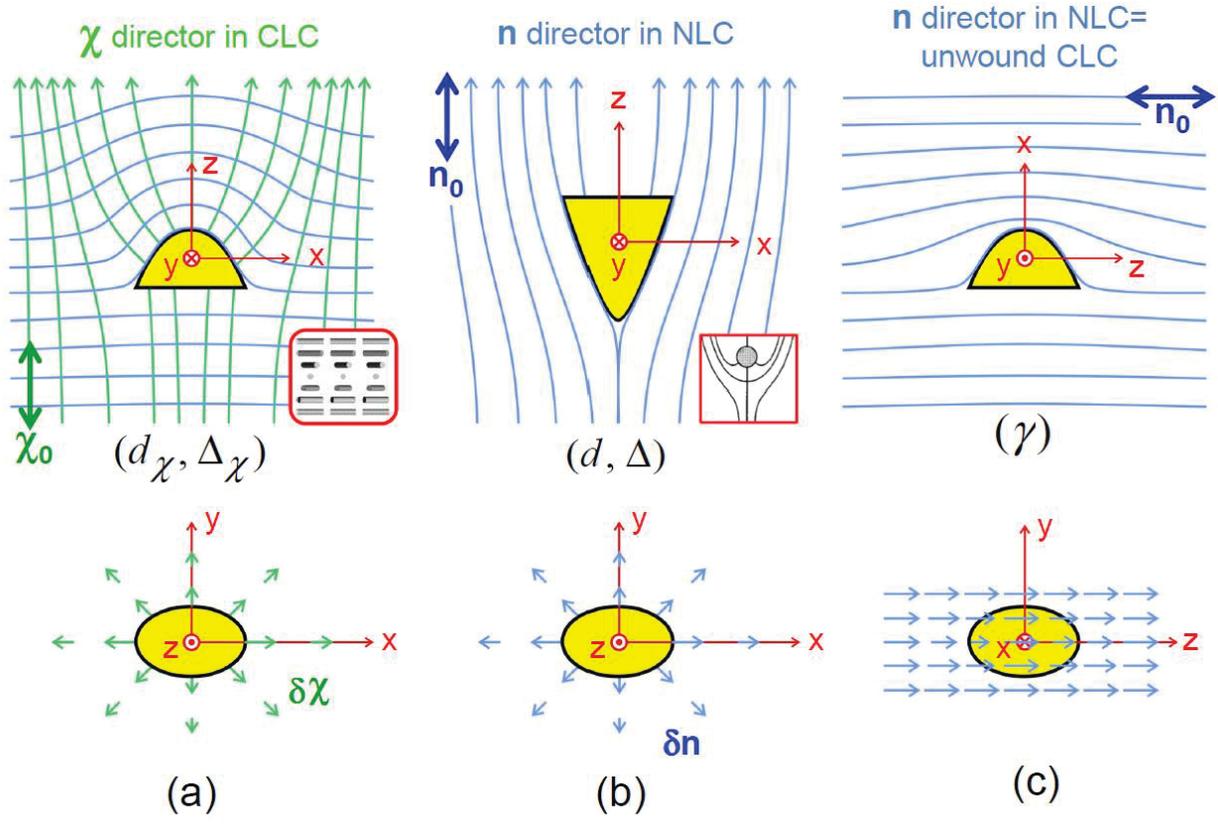

**Fig. S1**

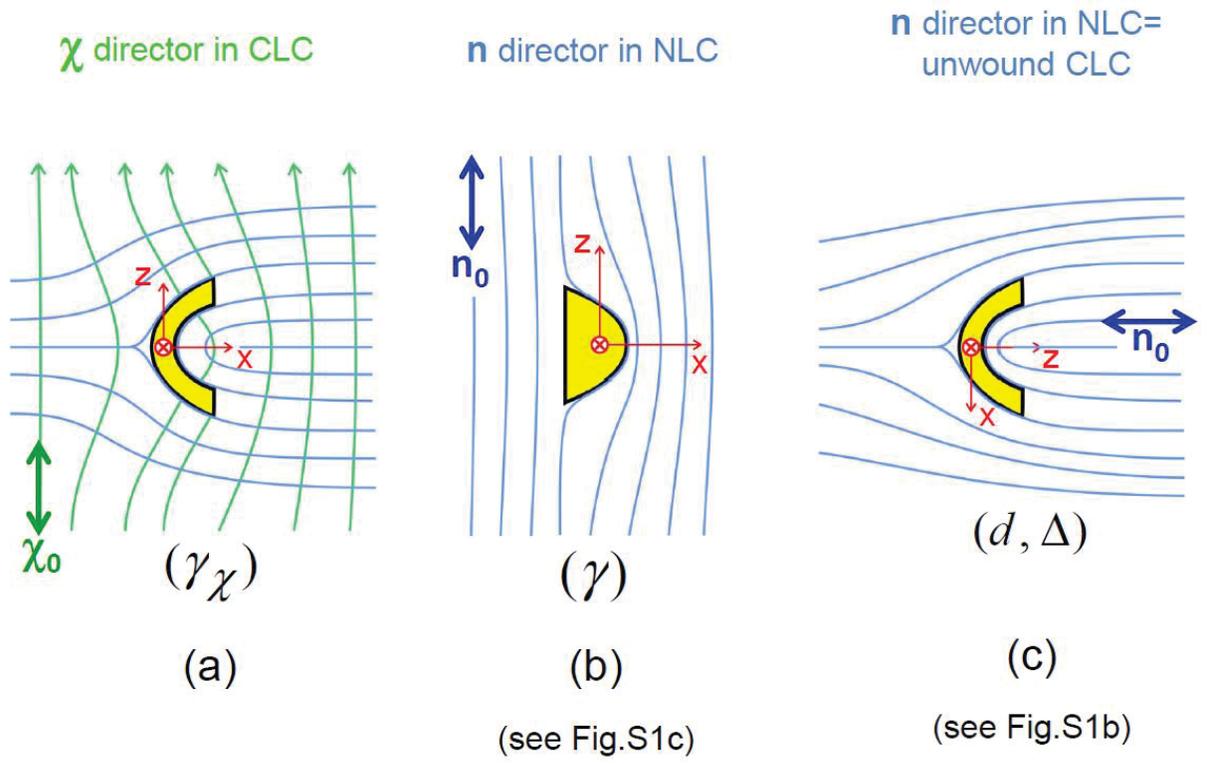

**Fig. S2**

(a)

(b) (see Fig.S1c)

(c) (see Fig.S1b)

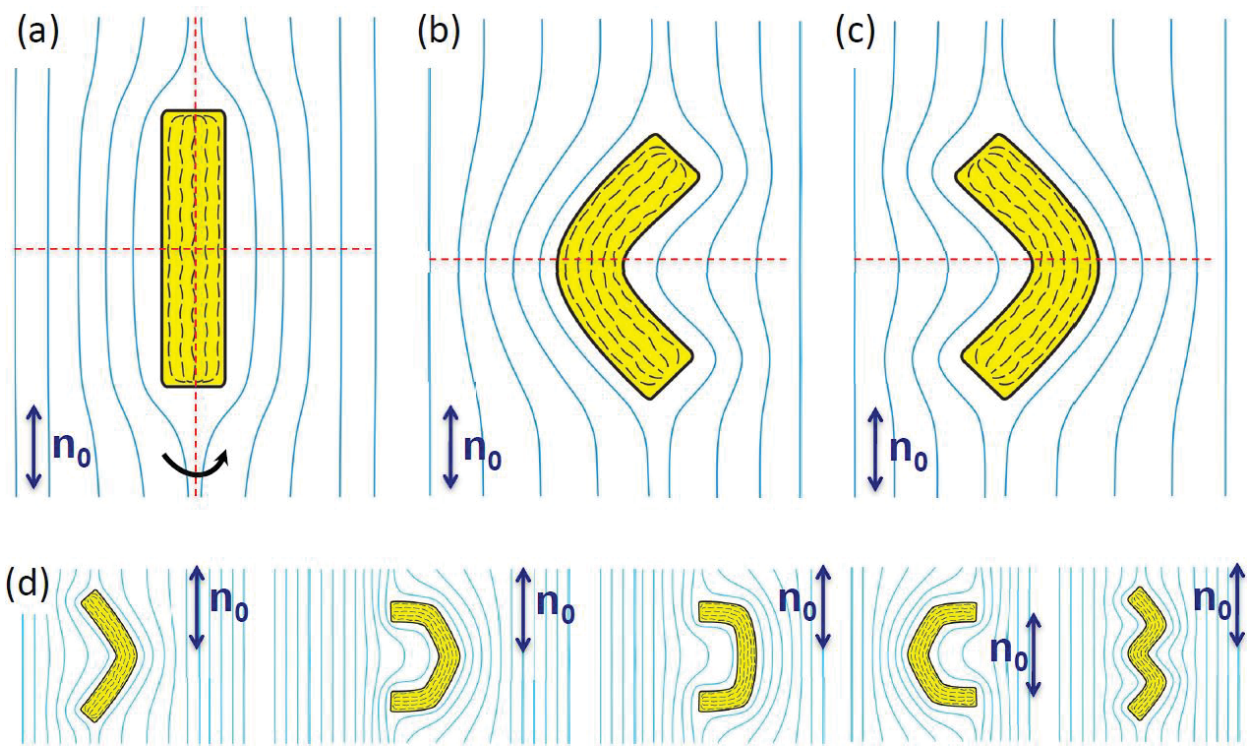

**Fig. S3**

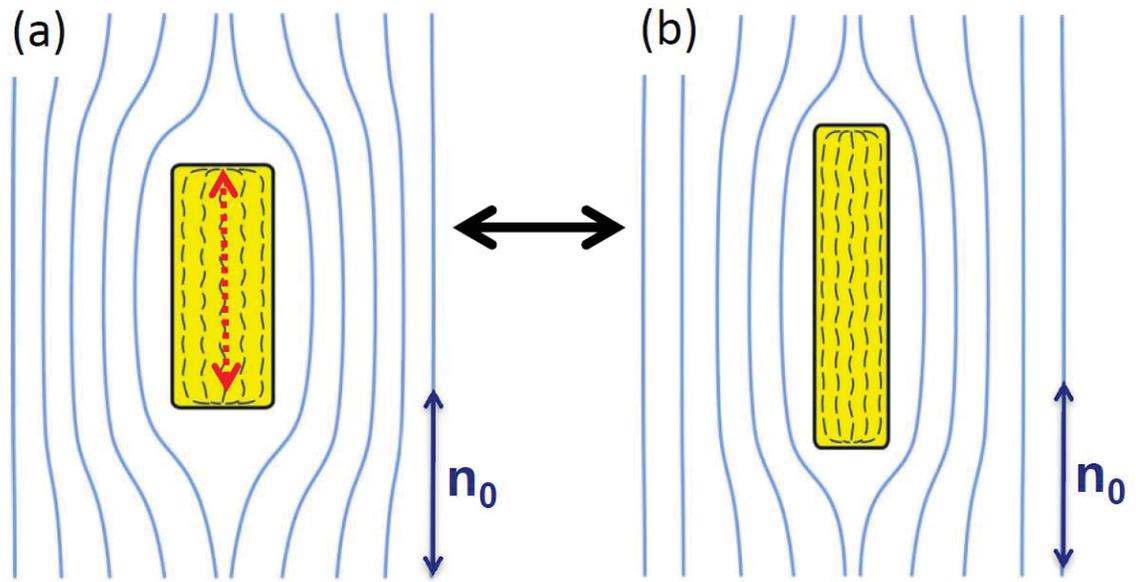

**Fig. S4**